# Mapping of Spin-Wave Transport in Thulium Iron Garnet Thin Films Using Diamond Quantum Microscopy

Rupak Timalsina,[1] Haohan Wang,[2] Bharat Giri,[2] Adam Erickson,[1] Xiaoshan Xu,[2] and Abdelghani Laraoui[1,2,*]

[1]Department of Mechanical & Materials Engineering, University of Nebraska-Lincoln, 900 N 16th St. W342 NH. Lincoln, NE 68588, USA

[2]Department of Physics and Astronomy and the Nebraska Center for Materials and Nanoscience, University of Nebraska-Lincoln, 855 N 16th St, Lincoln, Nebraska 68588, USA

*Email: alaraoui2@unl.edu

**Abstract**

Spin waves, collective dynamic magnetic excitations, offer crucial insights into magnetic material properties. Rare-earth iron garnets offer an ideal spin-wave (SW) platform with long propagation length, short wavelength, gigahertz frequency, and applicability to magnon spintronic platforms. Of particular interest, thulium iron garnet (TmIG) has attracted a huge interest recently due to its successful growth down to a few nanometers, observed topological Hall effect and spin orbit torque-induced switching effects. However, there is no direct spatial measurement of its SW properties. This work uses diamond nitrogen vacancy (NV) magnetometry in combination with SW electrical transmission spectroscopy to study SW transport properties in TmIG thin films. NV magnetometry allows probing spin waves at the sub-micrometer scale, seen by the amplification of the local microwave magnetic field due to the coupling of NV spin qubits with the stray magnetic field produced by the microwave-excited spin waves. By monitoring the NV spin resonances, the SW properties in TmIG thin films are measured as function of the applied magnetic field, including their amplitude, decay length (~ 50 μm), and wavelength (0.8 – 2 μm). These results pave the way for studying spin qubit-magnon interactions in rare-earth magnetic insulators, relevant to quantum magnonics applications.

## 1. Introduction

Magnetic insulators are of great interest for spintronics and magnonic platforms due to their low damping, allowing long-distance information transmission without Joule heating.[1] Yttrium iron garnet ($Y_3Fe_5O_{12}$, YIG), is a well-studied ferrimagnetic (FMI) insulator with a long spin-wave coherence length and relatively low damping ($2.3 \times 10^{-4}$), even for nanoscale spin-wave transmission lines.[2] When combined with heavy metals (HMs) like Pt, it exhibits spin transport phenomena such as spin Seebeck effect,[3] and enable transport of coherent spin waves due to spin orbit torque,[4] and spin transfer torque[5] effects within the Pt layer, extending its application to spintronic devices.[1] However, materials with perpendicular magnetic anisotropy (PMA) are desirable for such devices due to their potential for higher magnetic densities. While extensive work has been done on PMA ferromagnetic (FM) metals such as CoFeB,[6] insulators offer advantages over metals.[7] Their PMA originates from bulk resulting in low current densities when interfaced to HMs.[8–10] Rare-earth (RE) iron garnet insulators ($RE_3Fe_5O_{12}$, REIG) possess a complex unit cell with low damping and the dominant super-exchange interactions between magnetic sublattices $Fe^{+3}$ (Tetrahedral, Tet) with $RE^{+3}$ (Dodecahedral, Dod) and $Fe^{+3}$ (Octahedral, Oct) respectively gives rise the FMI behavior.[11,12] By selecting the RE element, the REIG magnetic properties (compensation temperature $T_M$, saturation magnetization $M_S$, magnetic anisotropy $K$, damping $\alpha$) can be tuned. In particular, thulium iron garnets ($Tm_3Fe_5O_{12}$:TmIG, **Fig. 1a**) with PMA have garnered significant interest due to the observation of spin orbit torque-induced switching effects.[13,14] The PMA originates from the combination of negative magnetostriction and in-plane compressive strain due to the lattice mismatch with $Gd_3Ga_5O_{12}$ (GGG) substrate,[13,14] leading to low current densities when interfaced with HMs for spin-orbit torque control. Notably, TmIG is also suggested to host topological spin textures,[8,15] further extending its applicability to high-density magnetic data recording.[16]

TmIG has been successfully grown as a thin film with thicknesses in the range of a few nanometers[17,18] and its magnetic damping constant was determined to be ~ $10^{-2}$ through frequency dependent ferromagnetic resonance (FMR) measurements.[19] Recent spin-wave (SW) electrical transmission studies on undoped[20] and Bi doped[21] TmIG films reported magnetostatic forward volume spin waves modes with a SW group velocity in the range of 0.15 – 4.9 km.s$^{-1}$, and a SW decay length in the range of 3.2 – 20.5 μm respectively. While these results are promising for using TmIG in magnonic spintronics,[1] there is no direct spatial measurement reported yet of its SW transport properties at the submicron scale.

Various techniques are used to study spatially dependent SW properties in magnetic films and nanostructures such as Brillouin Light Scattering (BLS)[22–24] and Magneto-Optical Kerr Effect (MOKE).[25–27] However, these techniques lack the combined spatial resolution (~ 300 nm for BLS and MOKE), experimental flexibility (variable temperatures and magnetic fields), and magnetic sensitivity required to image weakly magnetized materials such as two-dimensional (2D) magnets.[28] An alternative technique has recently emerged for nanoscale measurement of spin waves based on optical detection of the electron spin resonances of nitrogen vacancy (NV) centers in diamond.[29–33] Negatively charged NVs, composed of a substitutional nitrogen adjacent to a vacancy site, are bright and stable emitters that exhibit optically detected magnetic resonance (ODMR) and millisecond spin coherence at ambient conditions,[34] making them an ideal platform to investigate SW properties in REIG magnetic insulators. NV magnetometry has been utilized to probe magnetic dynamic excitations in ferromagnetic thin films and microstructures,[35–38] deduce the chemical potential of the SW bath in YIG,[30] and study the effect of propagating surface SW modes in YIG to amplify externally applied microwave (MW) to excite remote NV locations from the MW source.[29,39] Recently, by using electrically generated spin currents in Pt, an efficient tuning of the YIG magnetic damping was achieved and the amplitude of the magnetic dipole fields generated by a micrometer-sized resonant magnet enabled precise electrical control of the Rabi oscillation frequency of NV spins.[40] Indeed, NV magnetometry offers an enhanced sensitivity at the nanometer scale and has the capability to detect magnons with frequencies up to 100 GHz[41] and with variable wavevectors up to ~ $5 \times 10^7$ rad.m$^{-1}$.[42]

In this work we use NV magnetometry in combination with SW electrical transmission spectroscopy to study the properties (amplitude, group velocity, wavelength, and decay length) of microwave exited spin waves in 34 nm thick TmIG films grown on GGG substrates. We measure spatially resolved maps of the stray magnetic field produced by the microwave excited spin waves as a function of the amplitude of the applied magnetic field and find a SW wavelength of 0.8 – 2 μm, SW decay length of ~ 50 μm, much longer than earlier studies,[20,21] opening new opportunities of using TmIG in magnonic spintronics[1] and quantum magnonics.[43,44]

## 2. System, Fabrication, and Characterization

**2.1. TmIG thin film growth and characterization.** We used a pulsed laser deposition (PLD) to grow TmIG thin films (thickness of 34 ± 1 nm) on (111) GGG substrate, closely monitoring using in-situ reflection high energy electron diffraction (RHEED), **Fig. S1a**. As shown in **Fig. 1b**, the X-ray diffraction (XRD) spectra and the RHEED pattern (**Fig. S1a**) indicate an epitaxial growth. The Atomic force microscopy (AFM) topography map (**Fig. S1c**) confirms the smoothness of the grown film with a root mean square (RMS) surface roughness value ~ 0.25 ± 0.5 nm over a 20 μm × 20 μm scan. Moreover, the observation of the (222) XRD peaks indicate a lattice distortion of the TmIG films caused by the epitaxial strain, because the (222) diffraction peaks are forbidden for the bulk garnet body center cubic (bcc) crystal structure. The positions of the (222) peaks correspond to a 0.349 nm spacing between the (222) planes, which is smaller than the 0.355 nm bulk value, indicating an in-plane tensile strain. The thickness of fully strained film (critical thickness) is estimated as 5 nm from the width of the (222) peak, while the rest of the film is relaxed. The (444) TmIG diffraction peak is merged with GGG peak (**Fig 1b**) and it is clearly seen for a similar thickness TmIG film grown on (111) $Gd_{2.6}Ca_{0.4}Ga_{4.1}Mg_{0.25}Zr_{0.65}O_{12}$ (sGGG), **Fig. S1b**., explained by the different lattice constants of GGG (1.2382 nm) and sGGG (1.248 nm).[18]

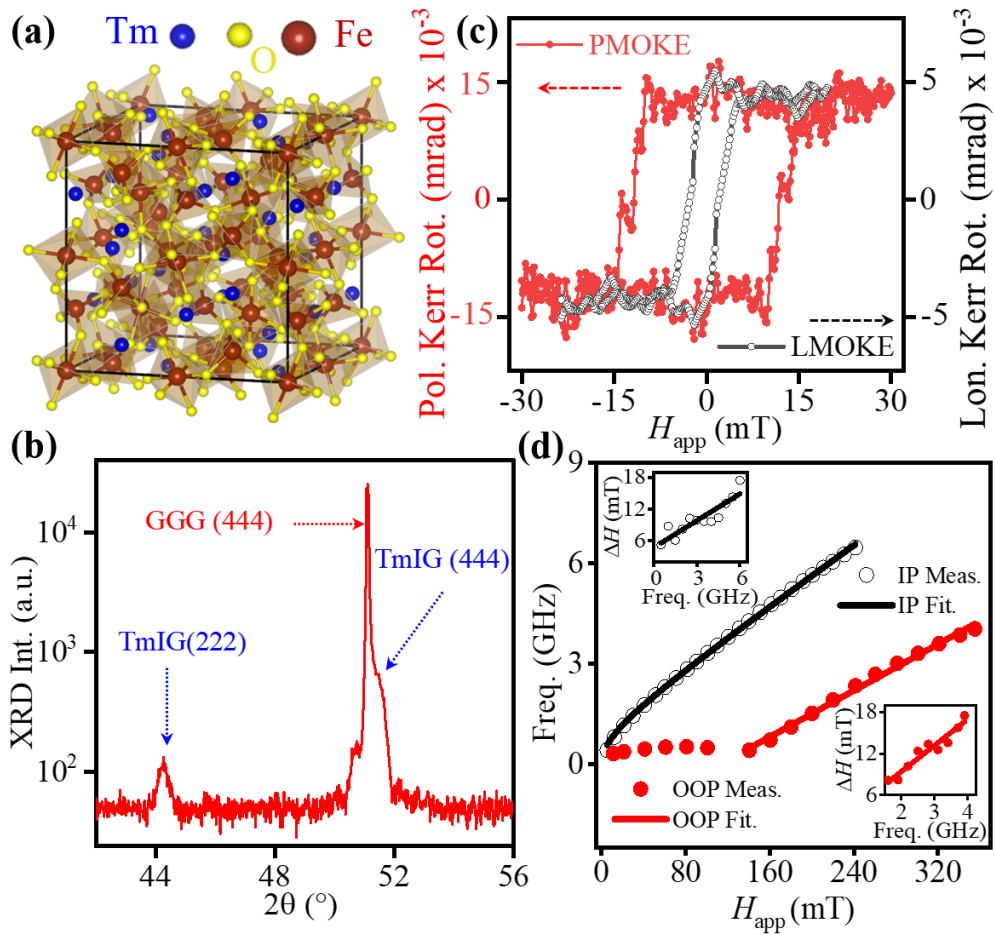

**Fig. 1:(a)** Crystal Structure of $Tm_3Fe_2(FeO_4)_3$ (TmIG). **(b)** Measured XRD spectrum of 34 nm thick TmIG film grown on 0.5 mm thick GGG substrate showing both TmIG and GGG peaks. **(c)** Polar MOKE (filled-circle-scattered line) and longitudinal MOKE (open-circle-scattered lines) curves *vs* applied magnetic field of TmIG (34 nm)/GGG. **(d)** In plane (IP, open circles) and out of plane (OOP, filled circles) FMR measurements on TmIG (34 nm)/GGG. Black and red solid lines are fits to the IP and OOP measurements respectively using the formulas in the main text. Insets of **(d)**: FMR linewidth of IP (open circles) and OOP (filled circles) FMR spectra *vs* MW frequency. The experimental values are fitted with black and red solid lines for IP and OOP FMR measurements respectively, described in the main text.

The magnetic properties of the 34 nm thick TmIG/GGG film have been examined using MOKE. As shown in **Fig. 1c** for polar MOKE (PMOKE, filled-circle-scattered line) and longitudinal MOKE (LMOKE, open-circle-scattered line) configurations, near-square-shaped hysteresis loops are observed with coercive field $H_C$ of 2.3 mT and 11.8 mT for LMOKE and PMOKE respectively. As shown in **Fig. 1c**, the polar Kerr rotation is ~ four times larger than the longitudinal values indicating a slightly out-of-plane (OOP) magnetic anisotropy. We performed vibrating-sample magnetometer (VSM) measurement (**Fig. S1d**) for in-plane (IP) configuration and found $H_C$ of ~ 0.2 mT lower than LMOKE, indicating some of the residual OOP magnetization in the LMOKE loop. The weaker PMA[45] in 34 nm thick TmIG/GGG appears to correlate with the smaller lattice mismatch between TmIG and GGG than that between TmIG and sGGG, where interesting effects such as interfacial Rashba-effect-induced anisotropy were observed.[18]

Ferromagnetic resonance (FMR) spectroscopy is used to measure the magnetic anisotropy, gyromagnetic ratio $\gamma$, and damping constant $\alpha$ values. **Fig. 1d** shows FMR frequency dependence with applied magnetic field $H_{app}$ curves for IP (open circles) and OOP (filled circles) configurations. The IP and OOP FMR resonance frequencies are determined from the derivative of real $S_{11}$ vs $H_{app}$ curves in **Fig. S3** (see the Supporting Information Section S3). The solid lines in **Fig. 1d** are fits based on the equations of IP and OOP measurements respectively:[18,20,21]

$f_{FMR,IP} = \mu_0 \gamma \sqrt{H_R(H_R + M_{eff})}; \quad f_{FMR,OOP} = \mu_0 \gamma (H_R + M_{eff})$, $H_R$ is the magnetic field corresponding the FMR resonance, $M_{eff}$ is the effective magnetization = $M_S - H_a$ for IP and = $M_S - H_\perp$ for OOP measurements. $H_a$ is the demagnetizing IP magnetic anisotropy, $H_\perp$ is the OOP uniaxial anisotropy.[46] The fitting parameters resulted in $\gamma = 22.4$ GHz/T, $H_a = 27$ mT, $H_\perp = 30$ mT, and $M_S$ of 66 kA/m. The $M_S$ value agrees well with the VSM measurement shown in **Fig. S1d**. We also plot the linewidth $\Delta H$ of the FMR IP and OOP measurements (insets of **Fig. 1d**) extracted from the FMR derivative curves in **Fig. S3** as a function of the MW frequency $f$. By fitting the $\Delta H$-$f$ curves with a $\mu_0 \Delta H = \mu_0 \Delta H_0 + \frac{4\pi}{\gamma \alpha f}$ (for IP) and $\mu_0 \Delta H = \mu_0 \Delta H_0 + \frac{2\alpha f}{\sqrt{3}\gamma}$ (for OOP), we find a damping value $\alpha$ of 0.0172 for IP measurements and 0.042 for OOP measurements respectively.

## 2.2. Spin-wave transmission spectroscopy of surface spin waves in TmIG thin films

For SW transmission spectroscopy, we employ an electrical detection method of spin waves using a vector network analyzer (VNA, Keysight model P5004A).[20,21,47] Ground-signal-ground (GSG) type antennas were fabricated on top of the TmIG film. Refer to **Fig. 2a, Fig. S2a** and the Supporting Information Section S2 for the nanofabrication details. These antennas are connected to the MW signal probes (**Fig. S2a**) which are then attached to the VNA using non-magnetic SMA cables. At a given applied MW frequency, a dominating SW mode is excited with a wavevector $k$, determined by the spatial GSG geometry of the stripline (**Fig. S2a**). We calculate the SW excitation spectra (**Fig. S2c**) by performing a Fast Fourier transform (FFT) of the stray in-plane field $B_y$ generated by the CPW GSG antenna along the $y$ direction (**Fig. S2b**). The calculated SW modes have wavevector values in the range of $k_1 = 0.2$ rad.$\mu m^{-1}$ for the main SW mode to > 5 rad.$\mu m^{-1}$ for the higher SW modes. In here, we focus mainly on measuring the propagating surface spin waves [22,39] for comparison with subsequent NV measurements.

**Fig. 2b** shows the in-plane real $S_{21}$ SW transmission spectral map as a function of $H_{app}$ applied along $x$ (parallel to the CPW). We plot the real $S_{21}$ spectra at two values of $H_{app}$ at 7 mT (**Fig. 2c**) and 15 mT (**Fig. 2d**), aligned along $x$ direction, leading to the excitation of magnetostatic surface spin waves (MSSW). A strong dominating mode just above the FMR frequency is seen with additional SW modes, highlighted by the dashed arrows in **Fig. 2c** and **Fig. 2d**. The intensity of the higher SW modes is weak in comparison to the main and the second SW modes, with good agreement to the calculated spectrum in **Fig. S2c**. To obtain the signal strength and phase information of the spin waves, we measured the imaginary part of $S_{21}$ parameter (Im$S_{21}$) as a function of $H_{app}$. The zoomed spectra are plotted in the insets of **Fig. 2c** and **Fig. 2d** corresponding to $H_{app}$ values of 7 mT and 15 mT respectively. We define $\delta f$ as the frequency difference between two adjacent maxima and minima of the Im$S_{21}$ (highlighted in the inset of **Fig. 2c**) corresponding to a SW phase of $\pi$. The SW intensity of the other SW peaks (modes) can be optimized by choosing the distance between the ground and source signal in the CPW to excite them efficiently.[47]

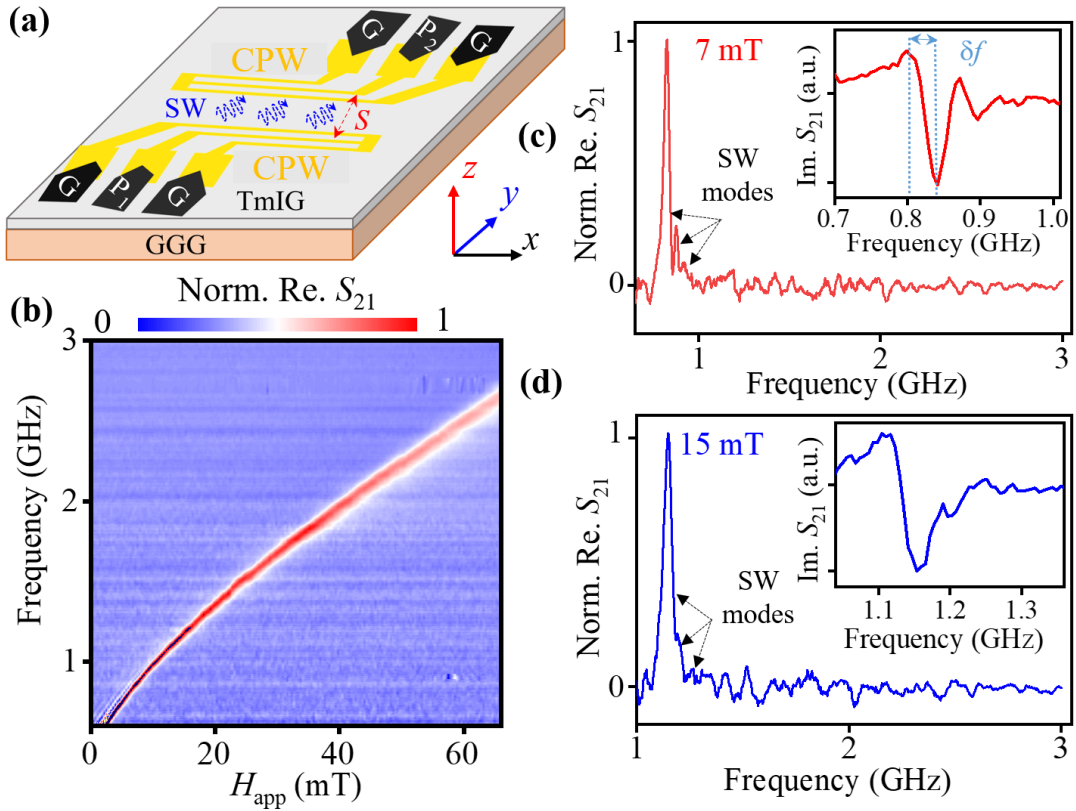

**Fig. 2:** (a) A schematic of the coplanar waveguides (CPWs) made on TmIG (34 nm)/GGG used for SW electrical transport measurement. (b) Normalized real $S_{21}$ intensity of the SW transmission map (MW frequency vs the applied magnetic field $H_{app}$) measured at a MW power of 0.1 mW. Normalized real $S_{21}$ intensity spectral profiles at selected magnetic fields of 7 mT (c) and 15 mT (d). Insets: Imaginary $S_{21}$ intensity spectral zoomed profiles measured at the same conditions as in (c) and (d). The SW modes are highlighted by the dashed arrows in (c) and (d).

The group velocity ($v_g$) of the spin waves is defined as: $v_g = 2\delta f \cdot S$, where $S$ represents the transmission distance of the propagating spin waves (~ 32 μm in **Fig. 2a**). We obtained $v_g$ in the range of 0.8 – 2 km/s. The spin-wave decay length ($l_d$) deduced from the equation $l_d = v_g/2\pi\alpha f$ falls in the range of 10 – 30 μm, comparable to CoFe thin films,[48] and higher than the values (~ 0.5 μm) obtained in 60 nm thick TmIG films with PMA grown on Nd$_3$Ga$_5$O$_{12}$ (NGG) substrates.[20] Spin-wavelength $\lambda$ values in the range of 0.8 – 2 μm are obtained depending on the magnitude of the applied magnetic field. However, it is quite difficult to obtain the exact values of $\lambda$ and $k$ from electrical measurements which are position-independent.[49] The MW power used for subsequent SW propagation is intentionally kept at 0.1 mW to avoid perturbations in the propagation due to magnetic nonlinearities.[50] To obtain accurate spatial measurements of the SW transport properties in the TmIG/GGG film, we use optically detected magnetic resonance imaging using NV magnetometry (discussed below).

## 2.3. NV magnetometry

The NV magnetic imaging platform is based on using an electronic grade (100) diamond substrate (2 mm × 1 mm × 0.08 mm) doped with a dense (~ 1 ppm) NV layer located ~ 6 – 10 nm beneath the surface. The diamond substrate with NV layer is placed face down on the TmIG film (**Fig. 3b**) to probe the interaction of spin waves with NV spin qubits.[31] We detail the diamond substrate and NV creation in the Supporting Information Section S4. For the experimental setup, we use an optically detected magnetic resonance (ODMR) microscope in the confocal geometry (schematic in **Fig. 3a**) to map surface spin waves in 34 nm thick TmIG/GGG film. A microscope objective (NA = 0.85) is used to focus a 532 nm laser (diffraction spot ~ 500 nm) and initialize the NV spin states in $|m_S = 0\rangle$. The resulting NV fluorescence (650-800 nm) is collected by the same objective through a dichroic mirror and detected by a single phone counting module (SPCM). Further details of the experimental setup are described in the Supporting Information Section S4. The NV spin-triplet ground state ($S = 1$) exhibits a zero-field splitting $D = 2.87$ GHz between states $|m_S = 0\rangle$ and $|m_S = \pm 1\rangle$.[34] Intersystem crossing to metastable singlet states takes place preferentially for NV centers in the $|m_S = \pm 1\rangle$ states, allowing optical readout of the spin state via spin-dependent fluorescence mechanisms.[34] The application of a bias magnetic field $H_{app}$ along the NV symmetry axis (111) breaks the degeneracy of the $|m_S = \pm 1\rangle$ state and results in two pairs of spin transitions for NV ensembles whose frequencies depend on the amplitude of magnetic field projection.[51,52] ODMR spectroscopy is implemented by sweeping the MW frequency via a gold (Au) wire (width = 10 μm, length of 200 μm, thickness = 100 nm) fabricated on the TmIG film (**Fig. 3b**), see the Supporting Information Section S2 for the fabrication details. When the MW frequency is resonant with NV spin transitions, there is a decrease of NV fluorescence. This reduction is related to the occurrence of intersystem crossing, which favors transitions to metastable singlet states, particularly in the $|m = \pm 1\rangle$ spin states. Four peaks appear in the ODMR map in **Fig. 3c** (on SiO$_2$/Si, no TmIG) performed at a distance of 5 μm from the Au wire at a MW power of 10 mW: $f_-$, $f_+$ for NV ensembles aligned along [111] and $f_{m-}$, $f_{m+}$ for NV ensemble merged and aligned on the opposite direction.[51,52] The bias magnetic field is applied 35° along the TmIG x$y$ plane and parallel to the Au wire, leading to almost in-plane saturation of the TmIG film.[53] In this geometry Damon–Eshbach spin-waves (DESWs) can be excited with the same MW injected through the Au wire. By tuning the direction and value of the magnetic field along the NV spin transitions one can probe spin waves with different wavelengths (wavevectors) discussed below.

## 3. Results and discussion

### 3.1 Effects of spin waves on NV spin properties

To investigate the effect of the propagating surface spin waves on the NV spins, we conducted identical measurement as shown in **Fig. 3d** where we obtain an ODMR map of the NV-doped diamond in contact with the TmIG film (**Fig. 3b**). A clear enhancement of the ODMR intensity of the low-transition frequency ($f_-$, $f_{m-}$, and $f_{m+}$) peaks is seen in **Fig. 3d**, explained by the frequency overlap between the stray microwave magnetic field generated by the spin waves and the NV spin transitions.[39]

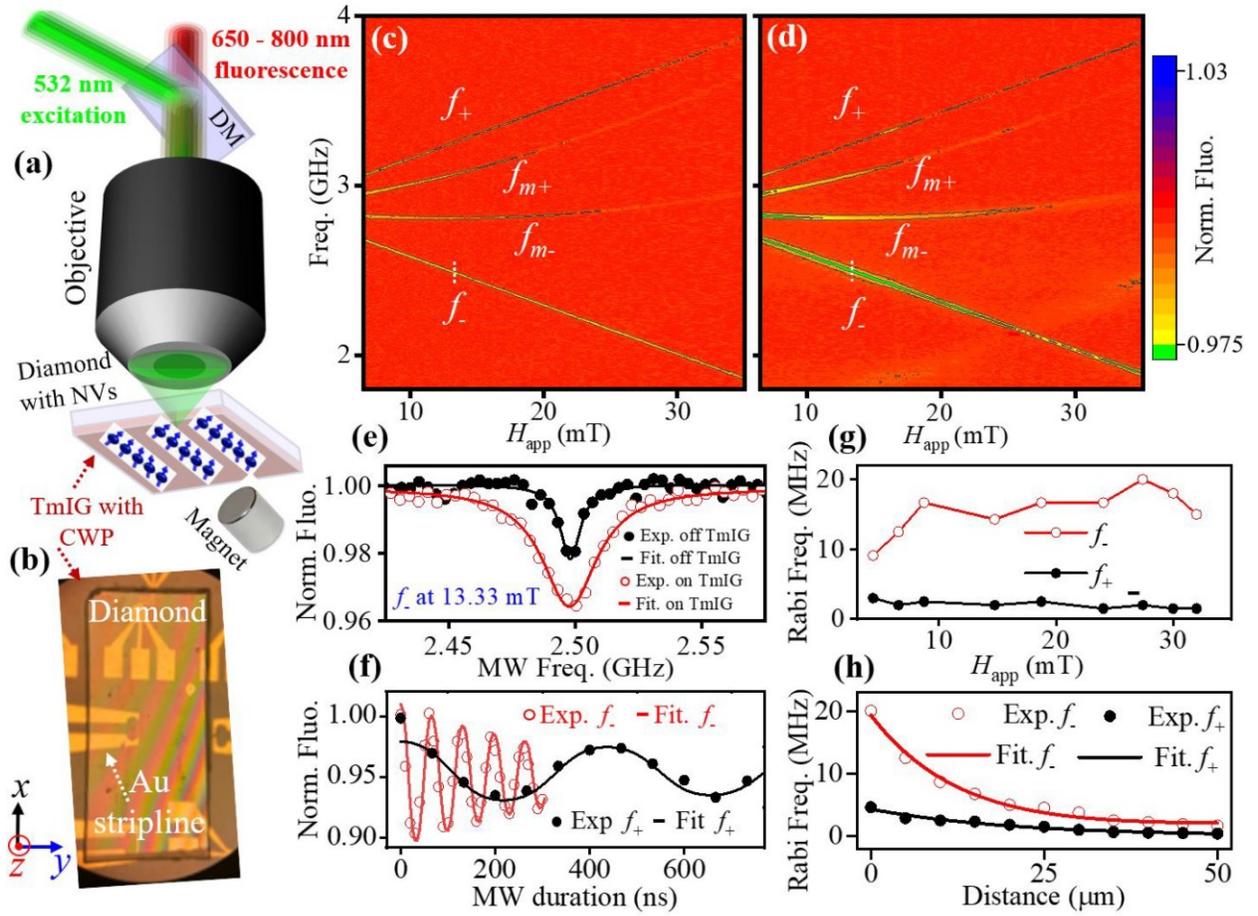

**Fig. 3:** (a) A schematic of the ODMR microscope in confocal geometry used to map SWs in TmIG/GGG. DM is a dichroic mirror. A green laser (532 nm) initializes the NV center spin and results in fluorescence in wavelength range of 650 − 800 nm collected by SPCM. The diamond face with doped NV layer is in contact with the TmIG film. A permanent magnet is aligned along the [111] axis of the (100) diamond to generate a magnetic field $H_{app}$, applied with an angle of 35° relative to the TmIG plane. (b) A picture of the diamond on top of TmIG with Au striplines to excite SWs for NV and $S_{21}$ measurements. NV ODMR frequency-$H_{app}$ map of diamond on top of Si (off TmIG), (c) and on TmIG (d) respectively. (e) ODMR $f_-$ signal recorded off TmIG (filled circles) and on TmIG (open circles) at $H_{app}$ = 13.33 mT, highlighted with dashed white lines in (c) and (d) respectively. The black and red solid lines are fits with Lorentzian function for off-TmIG and on-TmIG measurements respectively. A clear enhancement of ODMR contrast and linewidth is observed in the SW region of TmIG. (f) Rabi oscillations of $f_-$ peak (open circles) and $f_+$ (filled circles) recorded at $H_{app}$ = 13.33 mT and on top of the TmIG film. The MW power is 0.3 W. (g) Rabi frequency of $f_-$ peak (open circles) and $f_+$ (filled circles) measured as function of $H_{app}$. (h) Rabi frequency of $f_-$ (open circles) and $f_+$ (filled circles) peaks measured at $H_{app}$ = 13.33 mT as function of the distance from the stripline along the $y$ direction. The solid red (black) lines are the fitted curves for $f_-$ ($f_+$) measurements, see the main text for the fitting details.

Here, specifically we concentrate on the effect of the MW excited spin waves on the $f_-$ peak corresponding to the NV ensembles aligned along the NV axis ($\theta$ = 55° with $z$ axis). In **Fig. 3e** we plot the NV ODMR peak intensity for $f_-$ spin transitions, both off the TmIG film (filled circles) and on the TmIG film (open circles), fitted (solid lines) with Lorentzian function. A substantial enhancement of the ODMR contrast from ~ 2 % off TmIG to ~ 4% on TmIG. The ODMR linewidth $\Delta f$ varies from 9.62 MHz off TmIG to 26.57 MHz on TmIG respectively. These observations are attributed to the resonant energy exchange between the NVs and the propagating spin waves in the TmIG film.[31,54] Particularly, no effect is observed for the $f_+$ peaks at lower magnetic field, e.g. at 13.33 mT, explained by the lack of frequency overlap between the SW band and NVs spin transitions.[39] The SW amplitude can be deduced from NV magnetometry by measuring the MW magnetic field generated by the propagating spin waves as:[31]

$$B_{SW} = -B_{SW}^0 Re[(\hat{y} + isgn(k_y)\hat{z})e^{i(k_y y - ft/2\pi)}], \quad \text{Eq. 1}$$

$B_{SW}^0 = \mu_0 m_\perp^0 (1 + sgn(k_y)\eta)|\mathbf{k}|t_{TmIG}e^{-|k_y|d_{NV}}/2$, where $k_y$ is the wavevector of spin waves propagating along the $y$ direction, $f$ is the SW frequency, $m_\perp^0$ is the transverse magnetization corresponding to the magnetic stray field generated by the spin waves travelling perpendicularly to the static magnetization of the TmIG film (**Fig. 3a**), $\eta$ is the SW ellipticity, $d_{NV}$ is the NV-TmIG distance.

To measure accurately the MW magnetic field amplitude generated by the propagating spin waves we use Rabi spectroscopy[55] of NV spins in the proximity of the TmIG film. The NV Rabi frequency $\omega_{Rabi}$ can be retrieved by interfering the spin wave stray field with a reference MW field $B_{1,Ref}$ as:[31]

$$\omega_{Rabi}(y) = \sqrt{2}\gamma_{NV}\left|B_{SW}^0 \cos^2\left(\phi/2\right)e^{ik_y y} - B_{1,Ref}\right|, \quad \text{Eq. 2}$$

where $\gamma_{NV}$ is the NV electron spin gyromagnetic ratio (28 GHz/T), $\phi$ (~ 35°) is the angle between the TmIG film plane ($x$, $y$) and the NV axis. **Fig. 3f** displays the Rabi oscillation curves of $f_-$ (open circles) and $f_+$ (filled circles) at $H_{app}$ of 13.33 mT, fitted (solid lines) with a function $\sin(\omega_{Rabi} t)$ exp ($-t/T_{1,Rabi}$), $t$ is the MW duration. $T_{1,Rabi}$ (Rabi decay) is 2.27 µs for $f_-$ and 0.34 µs for $f_+$. The decrease of $T_{1,Rabi}$ of the lower $f_-$ spin transitions is explained by the presence of incoherent spin wave noise induced by the excitation of both microwave and thermal SW modes.[42] $\omega_{Rabi}$ is 2.27 MHz for $f_+$ and 15.38 MHz for $f_-$ spin transitions respectively. The increase of $\omega_{Rabi}$ of $f_-$ is further investigated as function of $H_{app}$ in the range of 4 – 32 mT (**Fig. 3g**). Frequencies up to 20 MHz are obtained for $f_-$ in the SW band region near (~ 0.5 µm) the Au stripline. For $f_+$, $\omega_{Rabi}$ is in the range of 1.5 – 3 MHz with a slight decrease at higher magnetic fields explained by the decrease of the MW transmission at higher frequencies of the MW amplifier used in the experiment (see the Supporting Information Section S4). $\omega_{Rabi}^{f_+}$ values correspond well the MW field induced by the Au stripline $B_1$ at the NVs' location ($y_{NV}$ = 5 µm and $d_{NV}$ = 0.8 µm). The distance $d_{NV}$ of ~ 0.8 µm between the NVs and TmIG film is estimated by optical, MW, and $T_1$ spectroscopy measurements (see the Supporting Information Section S5). **Fig. 3h** shows Rabi frequency $\omega_{Rabi}$ as function of the distance $y$ from the Au wire for $f_-$ and $f_+$ NV spin transitions at $H_{app}$ of 13.33 mT. $\omega_{Rabi}^{f_+}$ decreases away from the Au wire with a decay of ~ 20 µm, explained by the decrease of the $B_1$ field along the $y$ direction (see **Fig. S5a**). However, there is an amplification of $\omega_{Rabi}^{f_-}$ by one order of magnitude at low distances (< 10 µm) that decreases to ~ 3 MHz for distances up to 50 µm. $\omega_{Rabi}^{f_-}$ is fitted using equation **Eq. 2** including both $B_{1,Ref}$ and $B_{SW}^0$ effects. From the fitting parameters and SW amplitude (**Section 3.2.**) we get a spin-wave amplitude $m_\perp^0$ of 0.016 $M_S$ and a decay length $l_d$ ~ 50 ± 5 µm. To obtain the phase information of the spin waves, we employ ODMR continuous-wave imaging.

## 3.2. NV imaging of surface propagating spin waves in TmIG

We image the MW excited surface spin waves in 34 nm thick TmIG film by monitoring the NV $f_-$ ODMR intensity as function of the distance from the Au stripline (**Fig. 4a**). We apply a microwave (MW$_1$) to excite spin waves (above the FMR mode, **Fig. S6a**) for different values of $H_{app}$. To enhance the SW phase sensitivity of NV magnetometry we apply another microwave (MW$_2$) at the same MW frequency into a Cu wire (diameter of 25 μm) placed on top (~ 100 μm) of the diamond substrate and aligned perpendicularly to the Au wire (**Fig. 6a**). MW$_2$ field interferes with the stray-field induced by the propagating spin waves excited by MW$_1$ and leads to a spatial standing-wave pattern of the effective AC magnetic field which can be imaged using NV spins.[31,37,42]

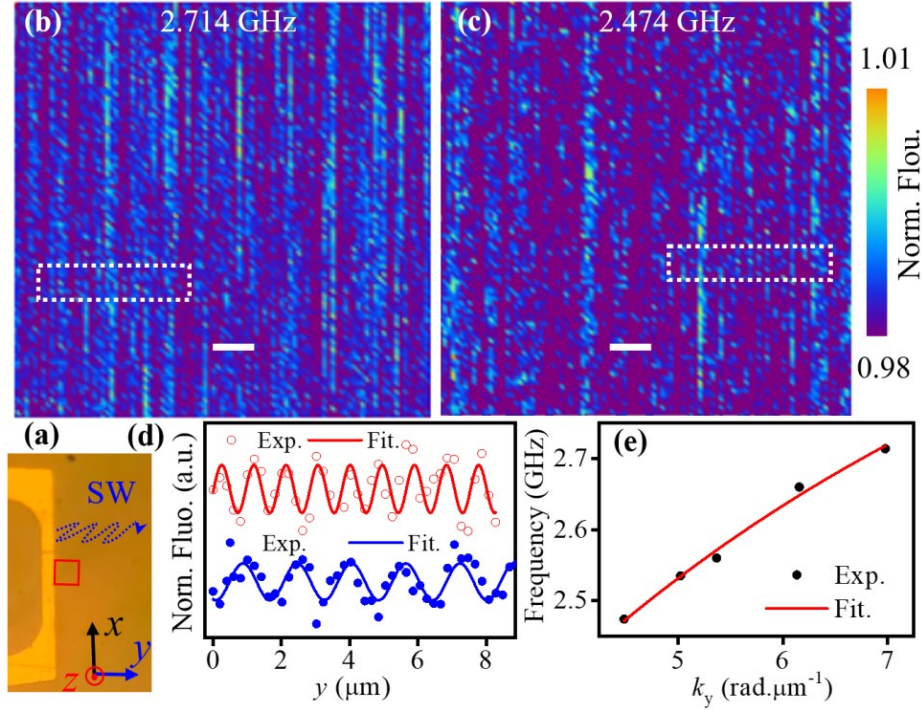

**Fig. 4:** *NV imaging of surface spin waves in TmIG film.* **(a)** a picture of the Au wire made on TmIG in contact with the NV-doped diamond substrate. Spatial maps of the normalized ODMR contrast while driving spin waves in TmIG at $f_-$ of 2.714 GHz at $H_{app}$ = 5.57 mT **(b)** and $f_-$ of 2.474 GHz at $H_{app}$ = 14 mT **(c)**. The mapped spatial region is highlighted by a solid red square in the inset of **(a)**. $P_{MW}$ = 10 mW and the scale bar in **(b)** and **(c)** is 2 μm. The ODMR contrast $FL/FL_0$ is obtained by normalizing the NV fluorescence with MW ($FL$) to that one without MW ($FL_0$) excitation. **(d)** Horizontal cross-sections of the ODMR maps integrated in the dashed rectangles in **b** (open circles), and **c** (filled circles), respectively, plotted vs distance $y$ from the Au wire. The solid lines in **(d)** are fits using **Eq. S4**. **(e)** Measured (filled circles) and calculated (solid line) DESW dispersion curve.

To calibrate the measurements on the TmIG/GGG film, we first imaged spin waves excited in 240 nm thick YIG/GGG film under similar conditions, **Fig. S6c**. A SW wavelength of ~ 10 μm is obtained on YIG, in a good agreement with earlier NV measurements.[31] In **Fig. 4b** and **Fig. 4c** we plot the 2D ($x$, $y$) NV normalized fluorescence intensity $FL/FL_0$ map at a frequency $f$ of 2.714 GHz ($H_{app}$ = 5.57 mT) and 2.474 GHz ($H_{app}$ = 14 mT) respectively. $FL$ ($FL_0$) is the fluorescence intensity with (without) MW excitation. **Fig. 4d** shows the integrated (over 10 lines) fluorescence intensity cross sections as function of $y$ distance in the dashed rectangles in **Fig. 4b** and **Fig. 4c**.

Clear oscillations are observed in the curves in **Fig. 4d** explained by the interference of the propagating spin waves excited by $MW_1$ with the uniform reference MW field ($MW_2$). The curves in **Fig. 4d** are fitted with the equation **Eq. S4** (see the Supporting Information Section S7) and the period of the oscillations corresponds to the wavelength $\lambda$ of the excited SWs. The FFT of the measured normalized fluorescence *vs* distance curves gives the wavevector $k_y$ value associated with the corresponding SW excitation MW frequency at given applied magnetic field. We detail the fitting and wavevector extraction in the Supporting Information Section S7. By plotting the frequency of the excited spin waves as function of the wavevector $k_y$ one can obtain the measured SW dispersion curve from the NV fluorescence spatial maps, **Fig. 4e**. The measured dispersion curve fits well with DESWs *f-k* dispersion curve using equation **Eq. S3** in Supporting Information Section S7. We find a good fit by using TmIG parameters ($M_S$ = 66 kA/m, $t_{TmIG}$ = 34 nm). $k_y$ increases from ~ 4.5 rad.µm$^{-1}$ at $H_{app}$ of 14 mT to ~ 7.5 rad.µm$^{-1}$ at $H_{app}$ of 5.5 mT.

To obtain the SW decay length $l_d$ in the TmIG film, we measure the 2D (*x, y*) normalized ODMR contrast map (**Fig. 5b**) in the area highlighted by the red rectangle in **Fig. 5a** (20 µm × 60 µm). By plotting the integrated normalized ODMR contrast over all *x* lines in **Fig. 5b** as function of the distance *y* from the Au wire, we extract a decay length $l_d$ of 50 ± 5 µm at $H_{app}$ of 12.03 mT (MW frequency is 2.533 GHz, and MW power is 10 mW). The high error in estimating $l_d$ in **Fig. 5c** comes from the spatial variation of the NV density across the diamond substrate,[51,52] that leads to high/low bumps in the NV fluorescence map (**Fig. 5b**). $l_d$ extracted from NV measurements (~ 50 µm) is higher than the value obtained by SW electrical transmission spectroscopy (10- 30 µm, **Section 2.2**.), that may be explained by the different excited SW modes. For example, $k_y \approx 0.2$ rad.µm$^{-1}$ for the dominating SW mode is SW transmission spectroscopy measurement (*i.e.*, SWs excited by CPW) and $k_y \approx 5$ rad.µm$^{-1}$ for NV measurements (SWs excited by the Au wire).

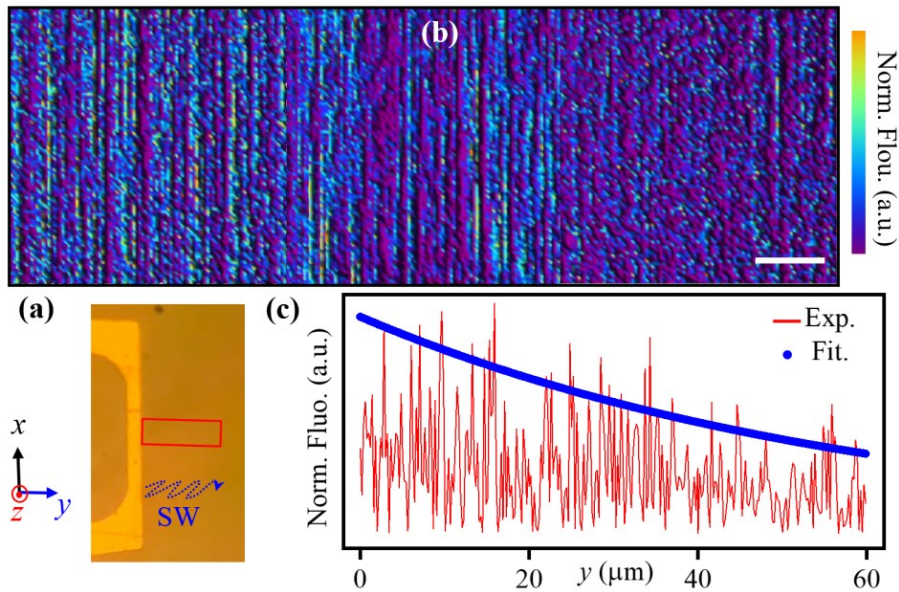

**Fig. 5.** *NV measurements of spin-wave decay length in 34 nm TmIG film.* **(a)** A picture of the Au stripline made on TmIG with the NV scanned area highlighted by the red rectangle. **(b)** Spatial map of the ODMR contrast while driving spin waves in TmIG at *f*− of 2.533 GHz at $H_{app}$ = 12.03 mT measured in the highlighted area (red rectangle) in **(a)**. The scale bar in **(b)** is 5 µm. **(c)** Integrated ODMR contrast over all lines along *x* as function of the distance *y* from the Au wire. The solid line in **(c)** is a decay fit of the SW amplitude using equation **Eq. 1** with a decay of 50 ± 5 µm.

### 3.3. Measuring interference of spin waves in TmIG thin film

To investigate spin-wave interference in the TmIG film we measure the 2D $(x, y)$ NV normalized fluorescence with both MW$_1$ and MW$_2$ applied on resonance with $f_-$ spin transitions at $H_{app}$ of 10.9 mT (MW frequency is 2.563 GHz and MW power of 10 mW) at the edge of the Au 200 μm long MW wire, **Fig. 6a**. At the center of the Au stripline we observe only one dominant spin-wave mode as in **Fig. 4** and **Fig. 5**, corresponding to the one dimensional case.[31] However, when measuring at the edge (inset of **Fig. 6a**) of the stripline we do see an interference pattern as shown in **Fig. 6b**. By performing vertical and transverse cross sections (integrated over 10 lines in the highlighted dashed lines in **Fig. 6b**) we see interference patterns for the vertical profiles in **Fig. 6c**. The second SW mode is excited by the MW passing through the edge wire (highlighted by a dashed white line) also known as edge effects, leading to obliquely propagating spin waves. The horizontal profiles in **Fig. 6d** show only one SW mode propagating perpendicularly along the MW stripline $k_y = 5.37$ rad.μm$^{-1}$.

Spin-wave interference is observed in 240 nm thick YIG/GGG film when using very weak MW powers,[33] *i.e.*, 500 times less than the MW power used in our experiments. The lower MW power regime prevents from broadening and saturating the NV ODMR peaks in the presence of spin waves (**Fig. 3e**), and therefore allows imaging of complex rich patterns of spin waves excited in different directions.[33]

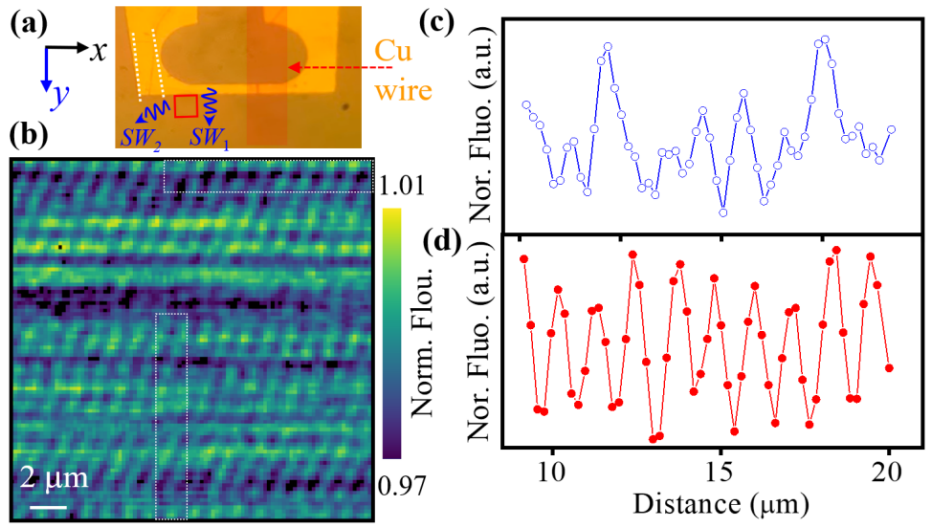

**Fig. 6:** *NV measurements of spin wave interference in 34 nm TmIG film.* **(a)** a picture of the Au stripline configuration on TmIG with diamond doped with NVs facing down. An additional Cu wire perpendicular to the Au stripline is used to induce SW interference with MW$_2$ reference $B_{1,Ref}$ field. **(b)** Spatial map of the ODMR contrast while driving spin waves in TmIG at $f_-$ of 2.563GHz at $H_{app}$ = 10.9 mT measured at the edge of the Au stripline. A clear SW interference is seen in **(b)** and it comes from the interference between two modes excited with different wavevectors from the center and the edge of the Au stripline. Integrated (over 10 lines) vertical **(c**, open-circle-scattered line**)** and horizontal **(d**, filled-circle-scattered line**)** cross-sections of the ODMR map in **(b)** plotted vs distance.

### 4. Conclusion

In summary, we use a dense layer of NV spin qubits in diamond to achieve precise phase sensitive magnetic resonance imaging of coherent surface spin waves in ferrimagnetic insulator TmIG films (thickness of 34 nm) epitaxially grown on GGG substrates. We measure spatially resolved maps of the stray magnetic field produced by the microwave excited spin waves and find a SW wavelength (0.8 – 2 µm), wavevector values in the range of 4 – 7 rad.µm$^{-1}$ that depends on the amplitude of the applied magnetic field and fits well with the dispersion curve of Damon–Eshbach surface spin waves. The spin waves last for distances up to 80 µm with a decay length $l_d \sim 50 \pm 5$ µm, much larger than the values found in earlier studies,[20,21] opening new applications for using TmIG in magnonic spintronics[1] and quantum magnonics.[43,44] By engineering the microwave striplines we observe a spin-wave interference between two single-mode SW modes excited by the microwave injected into the central and edge sides of the Au stripline. Studying the engineered spin-wave pattern may reveal an interesting physics of magnon interaction in ferrimagnetic insulators with PMA.[13,14]

This marks the first observation of propagating surface spin waves in TmIG thin films with low saturation magnetization ($M_S$ = 66 kA/m) as compared to YIG ($M_S$ = 142 kA/m). Our results demonstrate the capability of using NV magnetometry to effectively image spin waves (magnons) in weakly magnetized materials such as van der Waals 2D magnets.[28] Future measurements by using NV-diamond scanning probe microscopy (SPM)[56,57] could push down the spatial resolution from ~ 500 nm (confocal geometry) to below 50 nm (SPM) and provide more accuracy in filtering and imaging spin waves in ferrimagnetic insulators.[33] Particularly, TmIG with PMA is suggested to host topological spin textures[8,15] and therefore opens the door for NV magnetometry to study complex interaction between topological spin textures such as skyrmions and magnons.

**Supporting Information**
Supporting Information is available from the Wiley Online Library or from the authors.

**Acknowledgements**
This material is based upon work supported by the NSF/EPSCoR RII Track-1: Emergent Quantum Materials and Technologies (EQUATE) Award OIA-2044049 and NSF award# 2328822. The research was performed in part in the Nebraska Nanoscale Facility: National Nanotechnology Coordinated Infrastructure and the Nebraska Center for Materials and Nanoscience (and/or NERCF), which are supported by NSF under Award ECCS: 2025298, and the Nebraska Research Initiative. We thank K. Ambal for helping in setting up the FMR setup and B. Balasubramanian for initial help in VSM measurements.

**Author Contributions**
R.T. performed FMR, SW electrical transmission spectroscopy, and NV measurements; H.W. grew the TmIG/GGG films; H.W. and B.G. performed MOKE measurements; A.E. assisted R.T. in AFM measurements; X.X. and A.L. designed the experiments and supervised the project; A.L. wrote the manuscript with contributions of all authors.

**Conflict of Interest**
The authors declare no conflict of interest.

**Data Availability Statement**
The data that support the findings of this study are available from the corresponding author upon reasonable request.

## Supporting Information

### S.1. XRD analysis and AFM measurements

To check the epitaxial growth of thulium iron garnet (TmIG) on $Gd_3Ga_5O_{12}$ (GGG) and $Gd_{2.6}Ca_{0.4}Ga_{4.1}Mg_{0.25}Zr_{0.65}O_{12}$ (SGGG) substrates we used in-situ reflection high energy electron diffraction (RHEED, **Fig. S1a**) during the pulsed laser deposition (PLD) growth process. Additionally, we performed X-ray reflectivity (XRR) analysis (not shown here) to measure the thickness of the TmIG film and found $t_{TmIG} = \sim 34 \pm 1$ nm. We performed X-ray diffraction (XRD) spectroscopy at room temperature on TmIG/GGG substrate by using Rigaku SmartLab Diffractometer (Cu Kalpha radiation with a wavelength of $\sim 1.54°A$. **Fig. S1b** shows the measured XRD spectra recorded on 34 nm thick TmIG film grown on GGG (red curve) and sGGG (blue curve) respectively. The (444) TmIG diffraction peak is merged with GGG peak (**Fig. S1b**) and is visualized clearly in the TmIG film grown on (111) SGGG, **Fig. S1b**. To check the surface of the TmIG film, we used an atomic force microscopy (AFM, Bruker Innova AFM). **Fig. S1c** displays the AFM topography image of TmIG(34nm)/GGG. The AFM topography map shows a root mean square (RMS) surface roughness (RA) value of 0.289 nm over a 20 μm by 20 μm area.

To measure the bulk magnetic properties of the TmIG (34 nm)/GGG, we used vibrating-sample magnetometer (VSM, Versal Lab 3T from Quantum Design) and measured the in-plane (IP) film magnetization as function of the applied magnetic field at room temperature. The VSM measurements in **Fig. S1d** show a weak coercive field $H_C$ of $\sim 0.2$ mT lower than the longitudinal magneto-optical Kerr effect (LMOKE), indicating some of the residual of out-of-plane (OOP) magnetization in the LMOKE loop. The value of the IP saturation magnetization $M_S$ of $66 \pm 4$ kA/m corresponds well to the in-plane (IP) ferromagnetic resonance (FMR) measurements (see the main text).

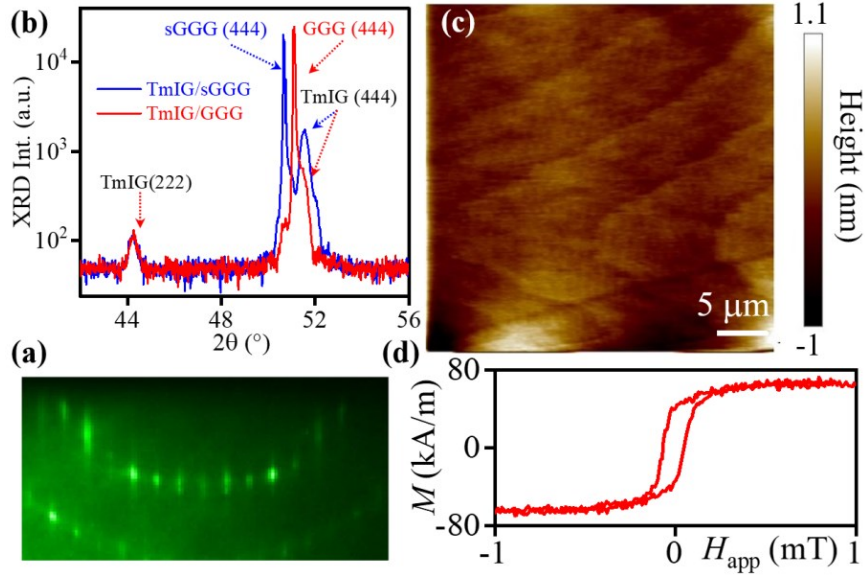

**Fig. S1:** **(a)** RHEED pattern of a 34-nm thick TmIG film grown on GGG. **(b)** Measured XRD spectra of 34 nm thick TmIG film grown on 0.5 mm thick GGG (red curve) and sGGG (blue curve) substrates showing TmIG, GGG, and sGGG peaks. TmIG (444) peak is clearly seen on sGGG substrate due to the high strain. **(c)** AFM image of 34 nm thick TmIG film grown on GGG, showing a smooth surface with an RA of 0.289 nm. **(d)** In-plane VSM $M$ vs $H_{app}$ hysteresis loop showing a low coercive field $\leq 0.5$ mT and a magnetization saturation $M_S$ of $66 \pm 4$ kA/m. $M_S$ value obtained from VSM **Fig. 1d** is used to get the anisotropy values using IP and OOP FMR measurements.

## S.2. Fabrication of the coplanar waveguide antennas and description of $S_{11}$ and $S_{21}$ measurements

The ground-signal-ground (GSG) coplanar waveguide (CPW) antennas are fabricated using Heidelberg Direct Laser Writing (DWL) 66FS Laser Lithography System. This system employs spatial light modulator-based surface structuring with a Directed Laser Beam source (wavelength = 405 nm, power = 60 mW). CPW lines are fabricated on top of TmIG (34 nm)/GGG with a 3.5 µm signal (S) and two ground (G) lines with separation distance of 5.8 µm between them to ensure a 50-ohm impedance match. Positive resist Microposit S 1813 is spin-coated on top of the TmIG film, followed by exposure using the DWL system and development using the developer AZ300-47. Subsequently, an $O_2$ plasma cleaning (50 mTorr, 50 sccm $O_2$, 100 W RF for 45 sec) process is performed to remove contaminants and residual photoresist after development using Trion Minilock- Phantom III Reactive Ion Etching (RIE) system. Finally, Ti (5 nm)/Au (100 nm) is deposited on the structure using AJA ATC-ORION 800 E-beam Evaporation System.

For FMR and spin-wave (SW) electrical transmission spectroscopy measurements, we used non-magnetic picoprobes (40A-GSG-150-DP) with GSG configuration and 150 µm separation between the ground and signal lines to inject the MW into TmIG (34 nm)/GGG (**Fig. S2a**). We measured $S_{11}$ parameters for the FMR measurements and measured the real and imaginary $S_{21}$ parameters for SW-transmission spectroscopy using Keysight Vector Network Analyzer (VNA, model P5004A). We used two large cylindrical permanent magnets (GMW Associates, M 3480) to provide a strong (up to 3 T) and uniform magnetic field for IP and OOP measurements. We measured the magnitude of the applied magnetic field $H_{app}$ using a Hall probe magnetometer. The MW power used to generate a MW magnetic field around CPW for the measurements is 0.1 mW.

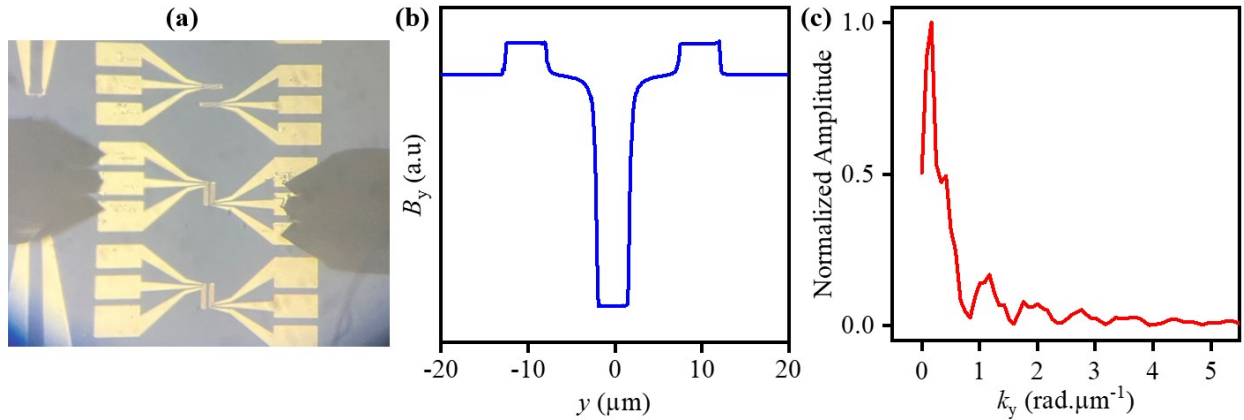

**Fig. S2:** (a) A picture of two non-magnetic picoprobes used to measure $S_{11}$ (for FMR) and $S_{21}$ (SW transmission spectroscopy), connected to the GSG CPW antennas. (b) The spatial profile of in-plane $y$ microwave field $B_y$ generated by the CPW. (c) Spin-wave excitation spectra obtained by Fast Fourier Transform (FFT) of $B_y$ in (b).

We employed "Electromagnetic waves frequency domain" finite element analysis solver and multiphysics simulation software COMSOL Multiphysics (https://www.comsol.com) to conduct simulations and predict the spatial profile of the stray magnetic field generated by MW injected into the GSG CPW. **Fig. S2b** shows the IP MW magnetic fields generated by the CPW structure along the $y$ axis by passing a microwave current through it. The Fast Fourier transform (FFT) provides simulated spin-wave excitation spectra of the 3.5 µm signal line with the main SW mode with a wave vector $K_1 = 0.2$ rad.µm$^{-1}$ and several other higher modes $K_2 - K_7$ leading to spin waves band with wave vectors up to 5 rad.µm$^{-1}$.

## S.3. FMR and spin-wave spectroscopy measurements

IP and OOP FMR measurements are done by measuring the $S_{11}$ parameter using VNA at room temperature. **Fig. S3** shows the FMR derivative spectra of a TmIG (34 nm)/GGG sample at a microwave power of 0.1 mW and at different MW frequencies (1, 2, 3, and 3.75 GHz) for IP **(a)** and OOP **(b)** configuration respectively. The IP and OOP FMR resonances occur at different magnetic fields, confirming in-plane magnetic anisotropy $H_a$ = 27 mT and OOP uniaxial anisotropy $H_\perp$ = 30 mT.[18] More details are provided in the main text in **Section 2.1**.

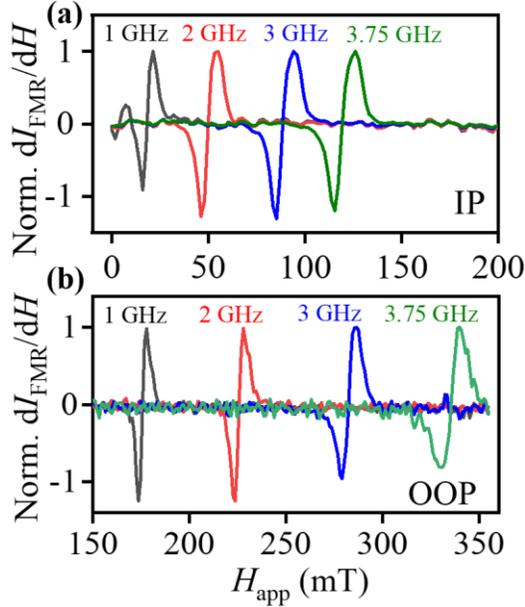

**Fig. S3: FMR measurements on TmIG (34 nm)/GGG film. (a)** Normalized in-plane (IP) FMR derivative spectra at 1 GHz, 2 GHz, 3 GHz, and 3.75 GHz as function of magnetic field $H_{app}$. **(b)** Normalized Out-Of-Plane (OOP) FMR derivative spectra at 1 GHz, 2 GHz, 3 GHz, and 3.75 GHz as function of the applied magnetic field $H_{app}$.

## S.4. Optically detected magnetic resonance setup.

The confocal microscope (CFM, **Fig. S4a**) used in this study is a home-built optically detected magnetic resonance (ODMR) setup combined with single photon counting module (SPCM) for confocal imaging.[58,59] We used a long working distance (5 mm), high NA (= 0.85) 100× Nikon objective to focus the green laser (532 nm) with a variable laser power (up to 1 mW). The diamond substrate with the NV layer facing down is attached to the TmIG with CPW antenna film by drop-casting an isopropyl alcohol (IPA) drop and letting it dry for a better adhesion.[31]

We used 2 mm × 2 mm × 0.5 mm type-IIa Electronic grade (100) diamond (Element 6) substrate with a nitrogen concentration < 5 ppb. The diamond is cut and polished along (100) plane at Delaware Diamond Knives Inc to 2 mm × 1 mm × 0.08 mm membranes. The diamond is then implanted at Cutting-edge Ions LLC with $^{15}N^+$ ions at an energy of 4 keV and a dose of 1. $10^{13}$ cm$^{-2}$ respectively to create a uniform layer of vacancies near the diamond surface (~ 6 – 10 nm).[60] We used Stopping and Range of Ions in Matter (SRIM) Monte Carlo simulations to estimate the vacancy distribution depth profile in the diamond substrate (**Fig. S4b**) and found a vacancy distribution within ~ 10 nm beneath the diamond surface facing the $^{15}N^+$ source. After the implantation we annealed the diamond substrate in an ultra-vacuum (pressure ≤ $10^{-6}$ torr) furnace at 1073 K for 4 hours and at 1373 K for 2 hours,[52] and then cleaned it for two hours in a 1:1:1 mixture of nitric, sulfuric, and perchloric acid at 473 K to remove graphite reside at the surface.[51,60,61] This process resulted in ~ 6 – 10 nm NV layer near the surface with a density of ~ 1 ppm, based on the fluorescence measurements.[59]

The NVs fluorescence (650 – 800 nm) is collected by the same objective, transmitted through the dichroic mirror, and sent to the detection setup to characterize the surface spin wave properties in TmIG using flip-mount mirrors. The fluorescence is then focused to a single mode fiber (L-com model 9/125, Single-mode Fiber APC Cable, FC / FC, 1.0 m) using 0.25 NA 20× objective and coupled to a single photon counting module (SPCM, Excelitas model # SPCM-AQRH-14-FC). A notch filter (Semrock, NF01-532U-25) is placed in the fluorescence path to block completely the 532-nm excitation light. The NV magnetic imaging of the TmIG film is performed by scanning the TmIG with diamond (travel range of 100 μm along $x$, $y$, and $z$) mounted on an open-loop three axis nanopositioning linear stage (NPXYZ100, Newport) and collecting the emitted photons from a confocal spot (~ 500 nm).

To monitor the position of the scanned diamond region from the CPW wire, we used a flip mirror to reflect and focus the NV fluorescence on a CCD camera (XIMEA USB 3 XiC MC031MG-SY) via a tube lens (Thorlabs, TTL200). For magnetic field alignment the NV fluorescence is reflected by another flip mirror and focused with a lens (focal length of 30 mm) into an avalanche photon detector (APD, Thorlabs APD410A), connected to a Yokogawa oscilloscope (DL9040L).

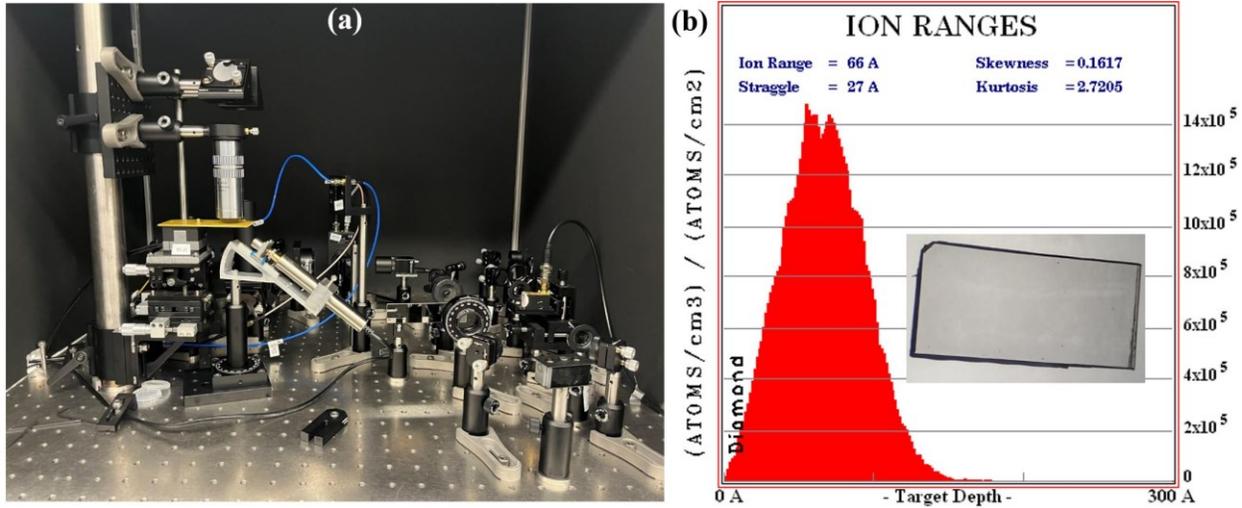

**Fig. S4: (a)** A picture of the ODMR confocal microscope integrated with an SPCM (not shown here). **(b)** SRIM vacancy-depth profile for the $^{15}N^+$ implantation in the diamond substrate (conditions are in the main text). Inset of **(b)** A picture of the diamond (2 mm × 1 mm × 0.08 mm) used in the ODMR measurements in this study. The implanted diamond side is facing TmIG film with CPW.

The MW is provided by two signal generators (SRS: Stanford Research Systems, SG384, and DS Instruments model SG12000PRO, 0.055 – 13 GHz), connected with two Mini Circuits switches (model ZASW-2-50DRA+). It is then connected to two MW amplifiers (Minicircuits ZHL-16W-43-s+ and ZVE-3W-83+) and sent through the blue SMA cables in **Fig. S4a**. The MW frequency is swept across the $|m_S = 0>$ to $|m_S = -1>$ ($f_-$) and $|m_S = 0>$ to $|m_S = +1>$ ($f_+$) NV spin transitions by sending a ramp function to the SRS generator analog frequency-modulation input. An output modulation ramp pulse is then sent to the oscilloscope and triggered with the APD (NV fluorescence) signal to perform live ODMR measurements. A permanent magnet (KJ magnetics, NdFeB N40) is mounted on a motorized Newport step motor (CONEX-TRA12CC) to provide a magnetic field in the range of 0 to 75 mT. There are four sub-ensembles of NV centers with different symmetry axes and in this study,[60,62] we used only the NV spins oriented along the [111] direction in (100) diamond for measuring the stray field generated by the propagating spin waves in the TmIG film. By monitoring the ODMR peaks in the oscilloscope along the [111] ($z$) direction, we can align the applied magnetic field very accurately. After the magnetic field alignment, the mirror is flipped down and the NVs fluorescence is detected by the SPCM.

## S.5. Estimation of NVs-to-TmIG distance

We used three approaches to estimate the NV to TmIG distance $d_{NV}$. We first calculated the MW induced magnetic stray-field for IP ($y$) and OOP ($z$) at a given MW power (*e.g.*, at 316 mW in **Fig. S5a**) by using the following Equation:

$$B_{str}(y,z) = \frac{\mu_0 I_{DC}}{2\pi w}\left(\arctan\left(\frac{wz}{y^2+z^2-\left(\frac{w}{2}\right)^2}\right)\hat{y} + \frac{1}{2}\ln\left(\frac{\left(y+\frac{w}{2}\right)^2+z^2}{\left(y-\frac{w}{2}\right)^2+z^2}\right)\hat{z}\right), \quad \text{Eq. S1}$$

where $w$ is the width of the Au wire, $I_{DC}$ is the DC current injected through the Au stripline. By measuring the Rabi frequency $\omega_{Rabi}$ (**Fig. 3f**) for the $f_+$ transition (no SW effect) one can estimate $d_{NV}$ to be ~ 0.745 ± 0.05 μm. We monitored the focus of the NV fluorescence in the CCD camera (inset of **Fig. S5a**) while shining the TmIG film with a white light source. We found ~ 1.2 μm focusing distance difference between the diamond (NV doped side) and the Au wire. The difficulty in estimating $d_{NV}$ with this technique is the limited $z$ spatial resolution of our confocal microscope (~ 1.2 μm). We then used NV $T_1$ relaxometry approach to determine accuracy of $d_{NV}$.

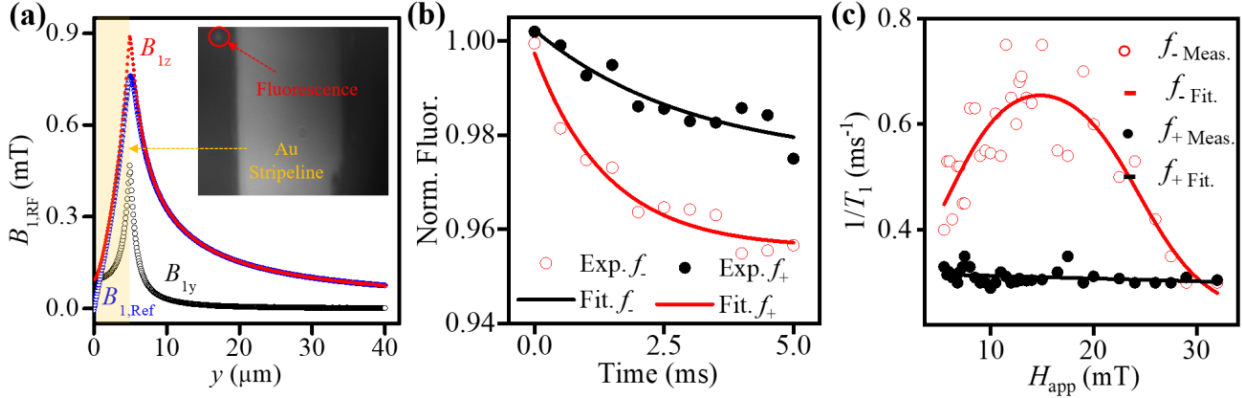

**Fig. S5: Estimation of the NV standoff from TmIG. (a)** Calculated magnetic stray-field along $y$ using **Eq. S1** (in plane) and $z$ (orthogonal) directions generated from injecting MW (power of 360 mW) on the Au wire (lateral size is 10 μm and thickness is 100 nm). Inset of **(a)** An optical image of the Au 10 μm wide stripline with NV fluorescence spot. **(b)** $T_1$ curve of NV spin transitions $f_-$ (open circles) and $f_+$ (filled circles) recorded at $H_{app}$ = 12.8 mT of diamond in contact with TmIG (34 nm) film and fitted with a single exponential decay function. **(c)** $1/T_1$ curve of NV spin transition $f_-$ (open circles) and $f_+$ (filled circles) recorded at $H_{app}$ = 12.8 mT. Solid red and black lines are fits for $1/T_1$ measurements for $f_-$ and $f_+$ spin transitions respectively using the same protocols as in reference 30.

The spin noise generated by spin waves is detected through a decrease of the NV spin relaxation $T_1$, which results in an overall reduction of fluorescence signal as function of time.[63] **Fig. S5b** shows NV-$T_1$ of $f_-$ (open circles) and $f_+$ (filled circles) NV spin transitions, fitted with solid lines (single exponential functions). The excited surface spin waves produce spin noise in the SW band and lead to a reduction of NV $T_1$ from 3 ms for $f_+$ to 1.4 ms for $f_-$ at 12.8 mT (**Fig. S5b**). The effect of spin waves is not seen for $f_+$ because there is no frequency overlap between the SW band and NV upper (higher frequency) spin transitions. The amplitude of the spin noise depends strongly on $d_{NV}$, TmIG film magnetic properties, and the applied magnetic field $H_{app}$. In **Fig. S5c** we plot the NV relaxation rate $1/T_1$ as function of $H_{app}$ for $f_-$ (open circles) and $f_+$ (filled circles) respectively. The fitted curves (solid red line for $f_-$ and solid black line for $f_+$) are based on these equations:[30,32]

$$\frac{1}{T_1^\pm} = C \frac{k_B T}{\hbar \omega_\pm} \int \left( \sin^4(\phi_k) + \cos^2(\theta) \frac{\sin^2(2\phi_k)}{4} + \sin^2(\theta)\sin^2(\phi_k) \right) \frac{\Delta H k e^{-2d_{NV}k}(1-e^{-2t_{TmIG}k})}{\pi\left(\Delta H^2 + [\omega(k,\phi_k) - \omega_\pm]^2\right)} k dk d\phi_k, \qquad \text{Eq. S2}$$

where $C$ is a constant, $k_B$ is Boltzmann constant, $T$ is temperature, $\omega_\pm$ are the NV frequencies for $f_+$ and $f_-$ spin transitions, $\theta$ (~ 55°) is the angle of the NV-axis with respect to the normal of the sample surface, $t_{TmIG}$ is the thickness of the TmIG film (34 nm), $\omega(k, \phi_k)$ is the SW dispersion function, $\phi_k$ is the angle between wavevector and IP projection of TmIG magnetization (~ 35°), and $\Delta H$ is the measured linewidth of the ferromagnetic resonance (**Fig. 1d**). The fitting for the measured $\frac{1}{T_1^\pm}$ vs $H_{app}$ curves gives a $d_{NV}$ of 0.8 ± 0.05 μm.

## S.6. Calibration of NV measurements on YIG/GGG

Before measuring on TmIG/GGG films we used similar SW imaging protocols as in references 33 and 37. We calculated (**Fig. S6b**) the ODMR peaks for $f_-$, $f_{m+}$, $f_{m+}$, and $f_+$ as a function of $H_{app}$. $f_{m+}$, $f_{m+}$ are the merged NV peaks in the opposite direction of [111]. In **Fig. S6b** we plot the ODMR peaks for diamond in contact with 240-nm thick YIG/GGG film (purchased commercially from Matsey Gmbh) at an applied magnetic field $H_{app}$ of 20.86 mT. There is an enhancement of NV ODMR for the lower frequency ($f_-$) spin transition, explained by the resonant energy exchange between magnetostatic surface spin waves (MSSW) and NVs.[39] To directly image MSSW we applied another MW, provided by a Cu wire (diameter of 25 μm) placed directly on top of diamond, and performed fluorescence imaging just above the FMR frequency of YIG (filled circles curve in **Fig. S6a**, MW power = 10 mW, MW frequency = 2.285 GHz at $H_{app}$ = 20.86 mT). By scanning the YIG film with diamond, we observe MSSW interference (**Fig. S6c**) with the MW field. The MSSWs spin wavelength is 10 μm, comparable to earlier measurements in YIG/GGG.[31] We then performed similar measurements on 34 nm thick TmIG/GGG film, discussed in the main text.

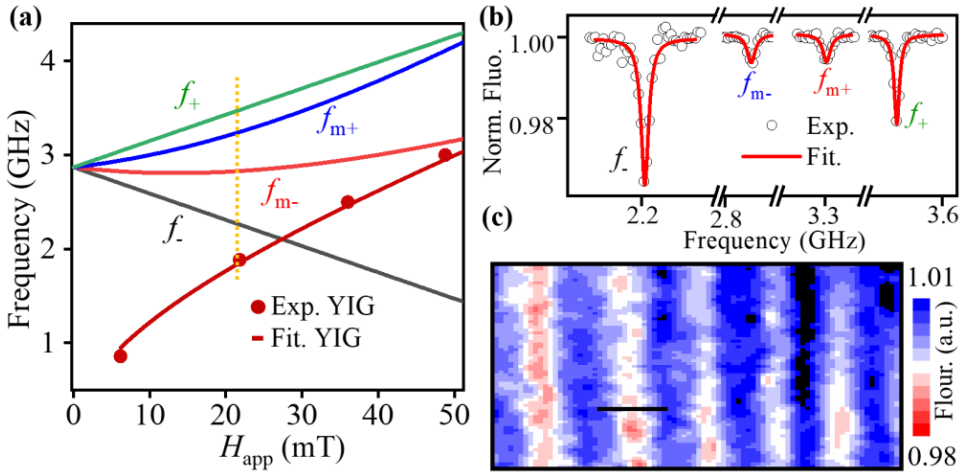

**Fig. S6**: **(a)** Calculated NV spin transitions (ODMR peaks) as function of $H_{app}$, overlapped with FMR IP curve of 240 nm YIG/GGG. **(b)** ODMR spectrum of NV ensemble aligned along [111] at $H_{app}$ of 20.86 mT. **(c)** NV ODMR contrast spatial map of the propagating MSSWs in in YIG (thickness is 240 nm). The scale bar in **(c)** is 10 μm.

## S.7. Dispersion relation and spin-wave analysis of NV-ODMR maps in TmIG

The SW dispersion $f$ vs $k$ relation describes the relationship between the frequency and wavevector of spin waves. In our setup configuration, we image Damon Eshbach Spin Waves (DESWs) and their frequency depends on the magnitude of the wavevector, the applied magnetic field, and the magnetic properties of the magnetic thin film using the following equation:[37]

$$f_{DESW} = \gamma \sqrt{\left(H_{app}\left(\sin(\theta_{NV} - \Delta\theta) + \left(\frac{M_S}{2}\right)^2\right) - \left(\frac{M_S}{2}\right)^2 \exp(-2k_y t_{TmIG})\right)}, \quad \text{Eq. S3}$$

Where $f_{DESW}$ is the frequency of DESWs, $\theta_{NV}$ and $\Delta\theta$ are the angle of NV axis to the surface normal ($\theta_{NV} \sim 55°$) and the canting angle of spins in the out of plane magnetic field direction ($\Delta\theta \sim 5°$) respectively.[37] $M_S$ is the saturation magnetization of TmIG film, $\gamma$ is the gyromagnetic ratio, and $t_{TMIG}$ is the thickness of TmIG thin film. We plot the calculated SW dispersion curve using Eq.3 ($M_S$ = 66 kA/m, $t_{TMIG}$ = 34 nm, $H_{app}$ = 7.5 mT) in **Fig. S7a**.

We used similar SW analysis to reference 33 to analyze the spatial maps shown in **Fig. 4** and **Fig. 5** in the main text and extract the wavevector of the imaged SW modes. In **Fig. S7b**, we plot the NV ODMR contrast map (normalized fluorescence intensity) at an applied magnetic field $H_{app}$ of 7.5 mT. We plot the integrated (10 lines) ODMR contrast horizontal cross section as function of the distance $y$ from the Au stripline, **Fig. S7c**. The measured curve is well fitted with the following equation:[33]

$$FL/FL_0 = 1 - \beta \frac{\omega_{Rabi}^2}{\omega_{Rabi}^2 + \delta^2}, \quad \text{Eq. S4}$$

Where $FL$ ($FL_0$) is the NV fluorescence with (without) MW, $\beta$ is a constant, and $\omega_{Rabi} = \frac{\gamma}{\sqrt{2}}(B_{SW} + B_{1,Ref})$ is the Rabi oscillation rate induced by the SW stray magnetic field ($B_{SW}$) and reference MW field ($B_{1,Ref}$). To extract the wavevector $k_y$ of the propagating DESWs, we plot (**Fig. S7d**) the corresponding FFT and found $k_y \sim 6$ rad.μm$^{-1}$.

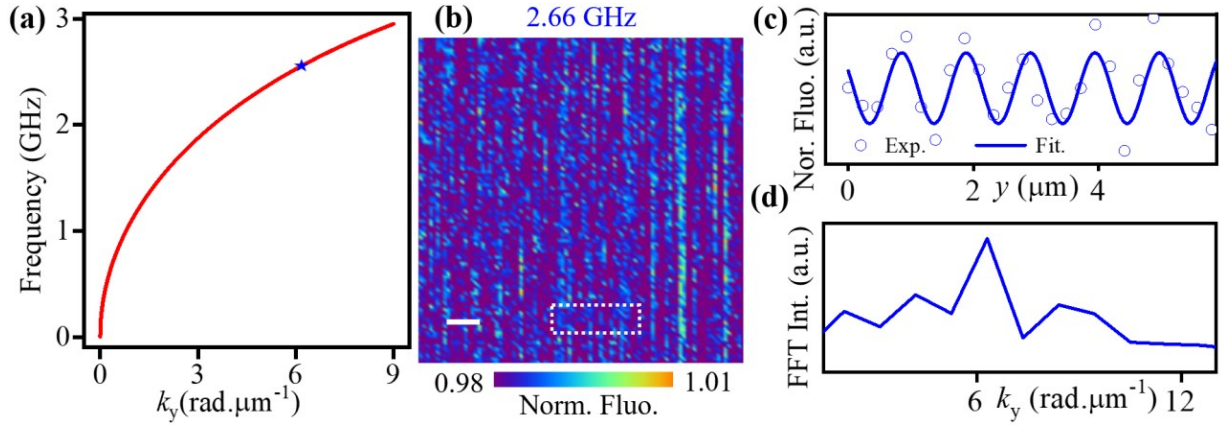

**Fig. S7:** (a) Calculated Dispersion curve for TmIG (34 nm)/GGG film at $H_{app}$ of 7.5mT. (b) Spatial map of the normalized ODMR contrast while driving spin waves in TmIG at $f_-$ of 2.66 GHz at $H_{app}$ = 7.5 mT. The scale bar in (b) is 2 μm. (c) Horizontal cross-section of the ODMR map integrated in the dashed rectangle in (b) plotted *vs* distance $y$ from the Au wire. (d) FFT spectrum of measured curve in (c) to get the wavevector of the propagating surface spin waves.